\documentclass[twocolumn,preprintnumbers,amsmath,amssymb]{revtex4}
\usepackage{graphicx}%
\usepackage{dcolumn}
\usepackage{amsmath}
\usepackage[dvips]{epsfig}

\makeatletter
\def\btt#1{\texttt{\@backslashchar#1}}%
\DeclareRobustCommand\bblash{\btt{\@backslashchar}}%
\makeatother

\begin{document}

\title{Orbital Order Instability and Orbital Excitations in Degenerate
       Itinerant Electron Systems}
\author{Feng Lu$^{1,2}$, Dong-meng Chen$^{1,2}$ and Liang-Jian Zou$^{1,
\footnote{Correspondence author, Electronic mail: zou@theory.issp.ac.cn}}$}
\affiliation{\it
 $^1$Key Laboratory of Materials Physics, Institute of Solid State Physics,
Chinese Academy of Sciences, P. O. Box 1129, Hefei 230031, China}
\affiliation{\it $^2$Graduate School of the Chinese Academy of Sciences}
\date{today}
\begin{abstract}
   We present the theory of orbital ordering in orbital-degenerate itinerant
 electron systems. After proposing the criterion of instability for orbital
 ordering or orbital density wave ordering, we find that the orbital and the
 spin-orbital collective excitation spectra in ferro-orbital ordered phase
 exhibit finite gaps. The anomalous electronic energy spectra manifested in
 the angle-resolved photoemission spectroscopy (ARPES) and
 the orbital occupation in the resonant X-ray  scattering (RXS) intensities
 are also presented for the orbital-ordered phase.
\end{abstract}

\pacs{75.10.Lp, 71.45.-d, 61.10.Eq}
\maketitle

     One of the most attracting developments in condensed matter physics
 this decade is to realize the roles of the orbital degree of freedom
 in the ground states, transport and optical properties in narrow-band
 transition-metal oxides and other compounds with unfulfilled $d$ or $f$
 orbitals. In these compounds, strong Coulomb interactions and highly
 anisotropic orbital correlations compel these nearly localized electrons
 to regularly occupy different orbitals, forming the orbital ordering (OO)
 \cite{Y.Tokura}.
 Upon doping from low to high concentration, the carriers, electrons or
 holes, are introduced into the compounds. One may expect that these carriers
 bring about frustrated orbital and spin arrangements, and eventually
 destroy the OO at a small doping concentration.
 However, it is found that in many such orbital compounds, the OO
 phase persists over wide doping range, as shown in the phase diagrams of
 $Ca_{2-x}Sr_{x}RuO_4${\ }\cite{M.Kubota, S.Nakatsuji},
 $(Sr,Ca)_{n+1}Ru_{n}O_{3n+1}${\ }\cite{G.Cao}, and $La_{1-x}(Sr,Ca)_{x
 }MnO_{3}${\ }\cite{Y.Tokura}, etc.
 This arises the question that why the correlated electrons are still OO
 in the presence of many carriers? Especially for the relatively wide
 band ruthenates, the radius of 4$d$ electrons is large
 and the 4$d$ electrons may exhibit itinerant character, in
 comparison with the localized 3$d$ electrons in manganites and vanadates.
 How do these 4$d$ electrons become localized over wide doped region?
 Understanding the formation of OO and the low temperature properties
 in the orbital-degenerate itinerant electron systems becomes the focus of
 this Letter, which will also help us understand the orbital properties in
 localized electron systems.

    As we know that the OO in orbital-degenerate narrow-band correlated
 electronic systems arises from the inter-orbital Coulomb repulsion U$'$
 \cite{Y.Tokura}.
 In an orbital-degenerate itinerant electron system, the inter-orbital
 Coulomb repulsion is usually not completely screened, then
 it may separate the two degenerate orbitals at a magnitude
 order of U$'$, and modify the orbital occupation at each site. Most
 probably, it is expected that the inter-orbital Coulomb interaction may
 play the similar role to the intra-orbital Coulomb repulsion, U,
 in itinerant electron systems \cite{Moriya}, and modulate the orbital
 distribution to form orbital density wave or OO phase.
 The conventional multi-orbital itinerant electron magnetism theory
 \cite{NORBERTO} ignored the possibilities of the formation of OO or
 orbital density wave phase. The recent experiment by Kubota
 et al. \cite{M.Kubota} showed that in relatively wide band 4d compound
 Ca$_{1.5}$Sr$_{0.5}$MnO$_{4}$ the metallic phase and OO coexist
 over the temperatures below 300 K, this fact is curious from the
 OO scenario in localized electron systems. These experimental and
 theoretical facts appeal for new theory for the orbital physics in
 itinerant electron systems.

  In this Letter we investigate the orbital physics in orbital-degenerate
 itinerant electron systems. We propose a criterion for the orbital
 instability for OO or orbital density wave ordering in the two-orbital
 itinerant electron model, and show that, similar to the Stoner condition
 in itinerant electron magnetism,
 the inter-orbital Coulomb interaction should be larger than a critical
 value for the occurrence of OO or orbital density wave state,
 independent of the fine structures of the degenerate bands;
 the energy spectra of the orbital wave and the spin-orbital wave exhibit
 finite gaps, which are distinguished different from the spin wave spectrum.
 Furthermore, we discuss the manifestations of the the electronic spectra
 in the angle-resolved photoemission spectroscopy (ARPES) and the orbital
 occupation resonant X-ray scattering intensity (RXS) in the ferro-orbital
 ordered itinerant electron systems.

 The Hamiltonian of two-orbital degenerate itinerant electron systems reads,
\begin{eqnarray}
  \hat{H}&=&\sum_{{\bf k},\sigma}\sum_{\alpha\beta}\epsilon_{{\bf k}\alpha\beta}
  \hat{c}_{{\bf k}\alpha\sigma}^{\dagger}\hat{c}_{{\bf k}\beta\sigma}
 +U\sum_{i{\alpha}}
  \hat{c}_{i\alpha\uparrow}^{\dagger}\hat{c}_{i\alpha\uparrow}
 \hat{c}_{i\alpha\downarrow}^{\dagger}\hat{c}_{i\alpha\downarrow}
  \nonumber\\
  &&+U'\sum_{i\sigma\sigma'}\hat{c}_{i1\sigma}^{\dagger}\hat{c}_{i1\sigma}
 \hat{c}_{i2\sigma'}^{\dagger}\hat{c}_{i2\sigma'}
  -J\sum_{i\sigma\sigma'}\hat{c}_{i1\sigma}^{\dagger}\hat{c}_{i1\sigma'}
  \hat{c}_{i2\sigma'}^{\dagger}\hat{c}_{i2\sigma}          \nonumber\\
  &&+J\sum_{i\alpha \neq \alpha'}
  \hat{c}_{i\alpha\uparrow}^{\dagger}\hat{c}_{i\alpha'\uparrow}
  \hat{c}_{i\alpha\downarrow}^{\dagger}\hat{c}_{i\alpha'\downarrow}
\end{eqnarray}
 The off-diagonal components of the non-interacting energy spectrum of electrons
 $\epsilon_{{\bf k}\alpha\beta}$ ($\alpha\neq\beta$) are usually finite,
 here for clarity, we consider only the diagonal hopping. We assume that
 the electron occupations of the two bands are the same to make sure that
 the system is para-orbital in the absence of interactions.
 In the present itinerant electron systems, considering the screening effect,
 we adopt the relation of the intra- and the inter-orbital Coulomb
 interactions $U$ and $U'$, and the Hund's coupling $J_{H}$:
 $U>U'>J_{H}>0$, rather than the relation: $U$=$U'$+$2 J$ \cite{prb18-4945}.

   As the response to the external field, when spin or orbital ordering
 occurs, the response function of the spins or that of the orbitals
 diverges near the disorder-order transition point. The spin and the
 orbital dynamical transverse susceptibilities \cite{Moriya} read,
\begin{eqnarray*}
  \chi_{s}^{-+}({\bf q},\omega+is)=i\int_{0}^{\infty}dt e^{i\omega t}
        <[S_{q}^{-}(t),S_{-q}^{+}(0)]>
\end{eqnarray*}
\begin{eqnarray}
  \chi_{o}^{-+}({\bf q},\omega+is)=i\int_{0}^{\infty}dt e^{i\omega t}
        <[\tau_{q}^{-}(t),\tau_{-q}^{+}(0)]>
\end{eqnarray}
 where the spin lift operator is $S_{q}^{+}$= $\sum_{{\bf k}m}
 \hat{c}_{{\bf k+q}m\uparrow}^{\dagger}\hat{c}_{{\bf k}m\downarrow}$
 and the orbital pseudospin lift operator is denoted as $\tau_{q}^{+}$=
 $\sum_{{\bf k}\sigma} \hat{c}_{{\bf k+q}1\sigma}^{\dagger}\hat{c}_{{\bf
 k}2\sigma}$.
 The spin and the orbital dynamical transverse susceptibilities in the
 paramagnetic and para-orbital metal regime are obtained in the random
 phase approximation.
 Although the full expressions of these susceptibilities are complicated,
 it is not different to find that in the present itinerant electron system,
 the spin instability for the spin density wave of wavevector {\bf q}
 occurs at:
\begin{eqnarray}
  -\frac{\eta_{1}^{'}({\bf q})+\eta_{2}^{'}({\bf q})}{\eta_{1}({\bf q})
    \eta_{2}^{'}({\bf q})+\eta_{2}({\bf q}) \eta_{1}^{'}({\bf q})}
\leqslant {U+J}
\end{eqnarray}
 with $\eta_{m}^{'}({\bf q})={1+(U-J)\eta_{m}({\bf q})}$ for the
 orbital index $m$, and
\begin{eqnarray*}
   \eta_{m}({\bf q})=\frac{1}{N}\sum_{\bf k}\frac{f_{{\bf k}+
   {\bf q},m\downarrow}-f_{{\bf k},m\uparrow}}{(\epsilon_{{\bf k}+
   {\bf q},m}-\epsilon_{{\bf k},m})+\Lambda_{m}}
\end{eqnarray*}
here
 $\Lambda_{m}=U(n_{m\uparrow}-n_{m\downarrow})+J_{H}(n_{\overline{m}\uparrow}
 -n_{{\overline{m}}\downarrow})$. In the paramagnetic phase,
 $n_{1\uparrow}+n_{2\uparrow}=n_{1\downarrow}+n_{2\downarrow}=n_{0}/2$.
 At $\Lambda_{m}=0$, $-\eta_{\alpha}(0)$ gives rise to the DOS per site
 for $\alpha$-band \cite{jpcm12-9985}. $n_{0}$ denotes the average
 occupation number of the electrons at each site.
 Restricting Eq.(3) to the single orbital case,
 $\eta_{2}(0)$ vanishes and $-\eta_{1}(0)$ gives rise to the
 DOS per site at Fermi energy $E_{F}$, the conventional
 Stoner condition for the single-band itinerant electron ferromagnetic
 order \cite{NORBERTO}: $U\varrho(E_{f}) \geqslant 1$, is thus
 unambiguously arrived.

 The orbital instability for the orbital density wave ordering of wavevector
 ${\bf q}$ occurs at:
\begin{eqnarray}
  \frac{\lambda_{\uparrow}^{'}({\bf q}) \lambda_{\downarrow}^{'}({\bf q})}
  {\lambda_{\uparrow}({\bf q}) \lambda_{\downarrow}^{'}({\bf q})+
  \lambda_{\downarrow}({\bf q}) \lambda_{\uparrow}^{'}({\bf q})}
  \geqslant {J}
\end{eqnarray}
 with $\lambda_{\sigma}^{'}({\bf q})={1+U'\lambda_{\sigma}({\bf q})}$, and
\begin{eqnarray*}
 \lambda_{\sigma}({\bf q})=\frac{1}{N}\sum_{\bf k} \frac{f_{{\bf k}
   +{\bf q},2\sigma}-f_{{\bf k},1\sigma}}{(\epsilon_{{\bf k}+{\bf q},2}
  -\epsilon_{{\bf k},1})+\Lambda_{\sigma}}
\end{eqnarray*}
 here $\Lambda_{\sigma}=(U-U')(n_{2\overline{\sigma}}-n_{1\overline{\sigma}})
 +(U'-J)(n_{1\sigma}-n_{2\sigma})$, and in the para-orbital phase,
 $n_{1\uparrow}+n_{1\downarrow}=n_{2\uparrow}+n_{2\downarrow}=n_{0}/2$.
 In the paramagnetic and para-orbital phase,
 $n_{1\uparrow}=n_{2\uparrow}=n_{1\downarrow}=n_{2\downarrow}$.
 $-\eta_{\alpha}(0)$ gives rise to the DOS per site for $\alpha$ band{\ }
 \cite{jpcm12-9985}. When the system is OO and paramagnetic, the condition,
 $n_{1\uparrow}=n_{2\uparrow}=n_{1\downarrow}=n_{2\downarrow}$, does not hold.
 This situation is ignored in the theory of the multi-orbital itinerant
 electron magnetism{\ }\cite{NORBERTO}.
 In the spin order and OO phase, the magnetization, $m=
 n_{\downarrow}-n_{\uparrow}$, and the orbitalization, $\tau=
 n_{2}-n_{1}$, are the order parameters describing the polarization degrees
 of spins and orbitals.

 We now focus on the ordered orbital ground state. The ordered phase can be
 ferro-orbital or orbital density wave of ${\bf q}$ when the ferro-orbital
 or the orbital density wave instability occurs:
\\
\noindent {\bf  A). Occurrence of Ferro-orbital Ordering}
\\
 In the paramagnetic and para-orbital itinerant electron systems with two
 degenerate bands, $f_{{\bf k},1\uparrow}=f_{{\bf k},1\downarrow}=
 f_{{\bf k},2\uparrow}=f_{{\bf k},2\downarrow}$,
 the divergence of the orbital dynamical transverse susceptibility at
 ${\bf q}=0$ gives rise to the criterion of ferro-orbital ordering
\begin{eqnarray}
   (U'-2{\ }J_{H}) \varrho_{12} \geqslant 1
\end{eqnarray}
 at $T=0 {\ }K$, where $\varrho_{12}=-\lambda_{1}({\bf 0})=-\lambda_{2}
 ({\bf 0})$ is the coalition DOS per site of each spin channel near the
 Fermi surface in the itinerant electron systems. Obviously the
 inter-orbital Coulomb repulsion $U'$ plays the same role in OO as the
 intra-orbital repulsion U in magnetism; while considering the Pauli
 exclusive principle and the Hund's coupling, the spins of the electrons
 tend to align in the same direction in different
 orbitals, so the Hund's coupling is unfavorable of the orbital
 polarization, making against to the emergence of the OO phase, as we see
 in Eq.(5).
 If the spins are complete polarized, the system is therefore half metallic,
 the condition for the ferro-orbital ordering instability becomes:
\[   ~~~~~~~~~~~~~~~~~
   (U'-J_{H})\varrho_{12} \geqslant 1 ~~~~~~~~~~~~~~~~~~~~~~~(5')
\]
 which implies that in the ferromagnetic phase, OO is more easier to occur
 than the paramagnetic phase. Since in the presence of the
 Hund's coupling and Coulomb interaction, due to the spin-orbital coupling
 in Eq.(1), the ordered spin structure favors the OO configuration.
\\

\noindent {\bf B). Formation of Orbital Density Wave State}
\\
 Due to itinerant character of the electrons, the orbital of the electrons
 may be not completely polarized as ferro-orbital ordering,
 rather, the orbital polarization may vary in different
 sites, it forms a spatial modulating structure with the characteristic
 wavevector ${\bf Q}$, i.e., orbital density wave phase.
 The criterion for the occurrence of orbital density wave phase is:
\begin{eqnarray}
       (U'-2{\ } J_{H})\varrho_{12}({\bf Q}) \geqslant 1
\end{eqnarray}
 in the paramagnetic regime, here $\varrho_{12}({\bf Q})=-\lambda_{1}({\bf Q})
 =-\lambda_{2}({\bf Q})$. In the ferromagnetic or spin density wave phase,
 Eq.(6) will be slightly modified.
 The conditions (5) and (6) show that at $T=0 {\ }K$, the orbital-degenerate
 itinerant electron system is ferro-orbital order or orbital
 density wave order when the inter-orbital Coulomb interaction strength $U'$
 is larger than a critical value to separate the two orbitals far enough.
%
%
%
\\

 \noindent {\bf C). Excitations in Ferro-Orbital Ordered Phase}
\\
    Furthermore, the dynamic instabilities of these response functions in
 Eq.(2) provide the information of collective and Stoner-type excitations
 of the spins and the orbitals.
 In the long wave-length and low frequency limit, we find that the
 energy spectra of the spin wave and the spin Stoner excitations in the
 present systems are consistent with the literatures \cite{NORBERTO}, and
 in analogous with the nondegenerate
 itinerant electron model with isotropic energy band \cite{Moriya}.
 Meanwhile, the external perturbation field also stimulates the orbital
 collective excitations, $i.e.$, the orbital wave in the long wave-length
 regime. And in the short wave-length regime, the orbital Stoner-type
 particle-hole excitations are also found. In the long wave-length and
 low frequency limit, the orbital wave spectrum $\omega_{o}({\bf q})$ is
 \begin{eqnarray}
     \Omega_{o}({\bf q})=\Delta_{o}+\xi_{o}{\ }{\bf q}^{2}
 \end{eqnarray}
 in the system with the
 spherical iso-energetic surface. Here the parameters $\Delta_{o}$ and
 $\xi_{o}$, depending on the band structure and the orbital polarization
 of the system, represent the gap and the stiffness of the orbital
 waves. Such kind orbital wave has been observed in narrow-band
 doped manganites by the Raman scattering experiments \cite{nature410-180}.
 And one also finds that the energy spectrum of orbital Stoner-type
 excitation, $\omega_{\sigma}$, is
 \begin{eqnarray}
   \omega_{\sigma}({\bf q}) = \epsilon_{{\bf k+q}1\sigma}-
          \epsilon_{{\bf k}2\sigma}
 \end{eqnarray}
 Since the spin angular momentum of the electrons is conserved during the
 hopping, one observes two independent branches in the orbital single-particle
 excitations (8).

 The difference of the dispersions of between the gapped orbital waves and
 the gapless spin waves could be understood in the framework of
 spontaneous symmetry breaking of the continuously symmetric systems.
 As we know that a remarkable feature of spontaneous symmetry breaking
 of the continuously symmetric order parameter is the appearance of
 massless particles: Goldstone bosons \cite{Y.Nambu,J.Goldstone},
 posses gapless excitations. Since the multi-orbital itinerant electronic
 Hamiltonian (1) is invariant under the continuous rotation of the spin
 operators in the spin space, so the spin wave is the Goldstone bosons
 without a gap in the energy spectrum. However, the continuous rotation
 invariance of the Hamiltonian (1) is broken in the orbital space, thus
 the orbital wave is not a type of Goldstone bosons, therefore the energy
 spectrum of the orbital waves exhibits a finite gap.

 Associated with the gapped orbital wave excitations, we find there exists
 a novel type of composite excitations, which arises from the interplay
 between spins and orbital degrees of freedom. According to the
 definition of spin-orbital composite operators \cite{prb61-6257}.
\begin{eqnarray*}
    \hat{K}_{\bf q}^{+}=\hat{c}_{{\bf k}+{\bf q}1\uparrow}^{\dagger}
         \hat{c}_{{\bf k}2\downarrow}+\hat{c}_{{\bf k}+
          {\bf q}2\uparrow}^{\dagger}\hat{c}_{{\bf k}1\downarrow}
\end{eqnarray*}
\begin{eqnarray}
    \hat{K}_{\bf q}^{-}=\hat{c}_{{\bf k}+{\bf q}1\downarrow}^{\dagger}
       \hat{c}_{{\bf k}2\uparrow}+\hat{c}_{{\bf k}+
        {\bf q}2\downarrow}^{\dagger}\hat{c}_{{\bf k}1\uparrow}
\end{eqnarray}
 the simultaneous spin-orbital dynamical transverse susceptibility becomes:
\begin{eqnarray}
    \chi^{-+}({\bf q},\omega+is)=i\int_{0}^{\infty}dt e^{i\omega t}
    <[\hat{K}_{\bf q}^{-}(t),\hat{K}_{\bf -q}^{+}(0)]>
\end{eqnarray}
 The energy spectrum of the simultaneous spin-orbital excitations can
 be extracted from the poles of the dynamical transverse susceptibility.
 In the long wave-length and low frequency regime, the energy spectrum
 of the collective excitations, or the simultaneous spin-orbital wave, is
\begin{eqnarray}
   \Omega_{so}({\bf q})=\Delta_{so}+\xi_{so}{\ }{\bf q}^{2}
\end{eqnarray}
 An energy gap $\Delta_{so}$ is also observed in the energy spectrum, and
 $\xi_{so}$ denotes the stiffness of the simultaneous spin-orbital wave.
 The corresponding Stoner-type particle-hole excitation spectrum is
\begin{eqnarray}
  \omega_{m}({\bf q}) = \epsilon_{{\bf k+q}m\uparrow}-
       \epsilon_{{\bf k}\bar{m}\downarrow}
\end{eqnarray}
 We expect that the simultaneous spin-orbital wave (11) and the orbital-dependent
 particle-hole excitations (8) and (12) could be identified
 in further resonant Raman experiments.

\begin{figure}[tp]
\vglue -0.6cm \scalebox{1.150}[1.15]
{\epsfig{figure=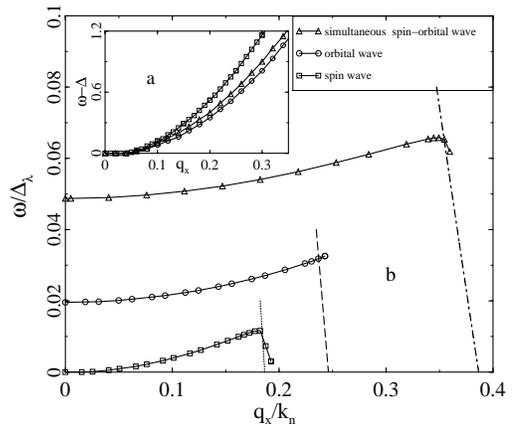,width=5.0cm,angle=270.0}}
\caption{Dispersion relations of the spin, the orbital and the
 spin-orbital collective and Stoner-type excitations along the (100)
 direction in ferro-orbital ordered phase.
 Dotted lines, long dashed lines and dot-dashed lines denote
 spin, orbital and simultaneous spin-orbital Stoner excitations,
 respectivity.
} \label{fig:fig1}
\end{figure}

 To understand the low-energy excitations of the ferro-orbital ordered
 itinerant electron systems, we numerically study the dependence of the
 energy spectra of the spin wave, the orbital wave and
 the simultaneous spin-orbital wave on the wavevector, and the
 corresponding Stoner-type particle-hole pair excitations, the more detailed
 dispersion relations of these excitations are shown in Fig 1.
 In the numerical calculation, we adopt the twofold degenerate parabolic
 energy bands, $\epsilon_{{\bf k}1}=$$\epsilon_{{\bf k}2}=$
 $(k_{x}^{2}+k_{y}^{2}+k_{z}^{2})/10$ eV, and the electron occupation is
 quarter-filling. The intra-orbital and inter-orbital Coulomb interactions,
 $U=1.55$ $eV$ and $U^{'}=1.44$ $eV$, respectively; and the Hund's coupling
 $J=0.055$ $eV$. The corresponding Stoner parameters, $k_{f}$ is 1.97, 1.61
 and 1.97, and $\Delta_{\lambda}$ is 0.13, 0.11 and 0.24 $eV$, respectivity.
 We find the z-components of the magnetization and the orbitalization
 per site, $m_{z}$ and $\tau_{z}$, are 0.4 and 0.3, respectively.
 In the long wavelength and low frequency limit, the gaps of
 the orbital-wave and the simultaneous spin-orbital wave are 2.25 and
 11.38 $meV$, respectively, which can be seen clearly in Fig.1.
 In the long wavelength limit, the dispersions of spin wave, orbital wave
 and the simultaneous spin-orbital wave with respect to the gaps are
 proportional to q$^{2}$, and the stiffness of the three kind collective
 excitations, as shown in the inset of Fig.1, are 13.65, 9.81 and 11.38 $meV$
 for the spin wave, the orbital wave and the simultaneous spin-orbital wave,
 respectively.

\begin{figure}[tp]
\vglue -0.6cm \scalebox{1.150}[1.15]
 {\epsfig{figure=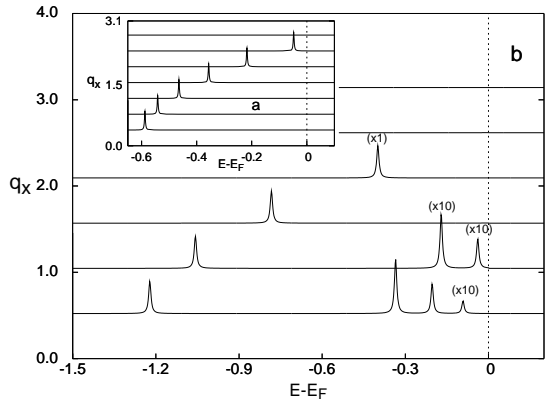,width=5.0cm,angle=270.0}}
 \caption{ Variation of electronic spectral function $A_({\bf k},\omega)$
 in paramagnetic and para-orbital phase (dot lines), and in ferromagnetic and
 ferro-orbital ordered phase (solid lines). The theoretical parameters are the
 same as Fig.1. Wavevector {\bf q} is along the x direction (100).
 }
\label{fig:fig2}
\end{figure}

 It is worthy of noticing that in the present orbital-degenerate itinerant
 electron system, the system remains metallic in the ordered orbital
 ground state. We find when the spin and orbital disorder-order transition
 occurs, the degenerate electronic energy spectrum splits into four different
 subbands in the random phase approximation.
 The variation of the electronic energy spectra near E$_{F}$
 can be reflected in the angle-resolved photoemission spectroscopy (ARPES).
 The spectral weight of itinerant electron in the OO and spin order phase is
 shown in Fig.2. When the system is para-magnetic and para-orbital,
 there is only one large peak in the intensity of the ARPES, which
 is contributed from the degenerate bands, and each band contributes the same
 weight to the peak, as shown in the inset in Fig.2.
 As far as the system enters into the ferro-orbital and ferromagnetic phase,
 the peak splits into four peaks, which correspond to the fact
 that the degenerate bands in orbital disorder phase splits into four
 different subbands since the degeneracy is lifted as the OO forms, hence
 the subbands with different weights contribute the four peaks with different
 height. So the measurement to the spectral functions by ARPES
 provides an indirect way to observe the orbital phase transitions,
 and the information of the variation of orbital polarization in multi-orbital
 itinerant electron systems.

\begin{figure}[tp]
\vglue -0.60cm \scalebox{1.150}[1.15]
{\epsfig{figure=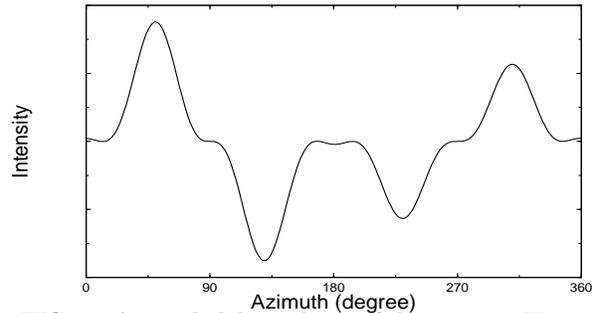,width=4.10cm,angle=270.0}}
 \caption{ Azimuthal dependence of the resonant X-ray scattering intensity
 in ferro-orbital itinerant electron systems.
 Theoretical parameters are the same as Fig.1.
 }
\label{fig:fig3}
\end{figure}

 The regularly polarized orbital occupation in narrow-band transition-metal
 oxides with OO can be manifested in the resonant X-ray scattering (RXS)
 experiment \cite{Y.Murakami}. In the present ferro-orbital itinerant electron
 system, according to the scattering theory of ferro-orbital ordering
 \cite{M.Kubota},
 we find the K-edge RXS intensity of 1s-4d event in the ferro-orbital itinerant
 electron system is significantly different from zero, and the amuzithal angle
 dependence of the RXS intensity is shown in Fig.3. The period of the RXS
 intensity is 2$\pi$, and the azimuthal dependence of the intensity is very similar
 to that in ferro-orbital ordered Ca$_{1.5}$Sr$_{0.5}$RuO$_{4}$.
 However we notice that ruthenates are almsot three-orbital degenerate, the
 two-orbital degenerate itinerant electron compound is not found yet.
 We expect more extensive search for such systems in experiments to examine
 the present itinerant orbital order theory.
\\
\\

   This work was supported by the NSFC of China and the BaiRen Project of
 the Chinese Academy of Sciences (CAS). Part of numerical calculation was
 performed in CCS, HFCAS.

\maketitle
\bibliography{apssamp}

\end{document}